\documentclass[twocolumn,pre,numerical,superscriptaddress,longbibliography]{revtex4-1}
\usepackage{graphicx,color,epsfig,subfig,dcolumn,bm,mathrsfs,amsmath,amssymb}

\newcommand{\Pe}{\mathrm{Pe}}

\newcommand{\change}[1]{\textcolor{black}{#1}}

\begin{document}

\title{Cross-interaction drives stratification in drying film of binary colloidal mixtures}

\author{Jiajia Zhou}
\email[]{jjzhou@buaa.edu.cn}
\affiliation{Key Laboratory of Bio-inspired Smart Interfacial Science and Technology of Ministry of Education, School of Chemistry and Environment, Beihang University, Beijing 100191, China}
\affiliation{Center of Soft Matter Physics and Its Applications, Beihang University, Beijing 100191, China}

\author{Ying Jiang}
\email[]{yjiang@buaa.edu.cn}
\affiliation{Key Laboratory of Bio-inspired Smart Interfacial Science and Technology of Ministry of Education, School of Chemistry and Environment, Beihang University, Beijing 100191, China}
\affiliation{Center of Soft Matter Physics and Its Applications, Beihang University, Beijing 100191, China}

\author{Masao Doi}
\email[]{masao.doi@buaa.edu.cn}
\affiliation{Center of Soft Matter Physics and Its Applications, Beihang University, Beijing 100191, China}


\begin{abstract}
When a liquid film of colloidal solution consisting of particles of different sizes is dried on a substrate, the colloids often stratify, where smaller colloids are laid upon larger colloids. 
This phenomenon is counter intuitive because larger colloids which have smaller diffusion constant are expected to remain near the surface during the drying process, leaving the layer of larger colloids on top of smaller colloids. 
Here we show that the phenomenon is caused by the interaction between the colloids, and can be explained by the diffusion model which accounts for the interaction between the colloids.  
By studying the evolution equation both numerically and analytically, we derive the condition at which the stratified structures are obtained.
\end{abstract}


\maketitle


Drying of a colloidal film is important in many places such as in printing \cite{Tekin2008}, spreading and coating \cite{Taylor2011} and material science \cite{Juillerat2006, Keddie_Routh}. 
An important problem is how the structure of dried film is controlled by drying conditions.  
It is known that the spatial distribution of colloidal particles in the drying process is determined by two competing processes. 
One is the Brownian motion \cite{RSS, Dhont, DoiSoft} which is characterized by the diffusion constant $D$, and the other is evaporation \cite{Keddie_Routh}, characterized by the speed $v_{\rm ev}$ at which the surface recedes.  
The competition between them can be quantified by the film formation Peclet number $\Pe = v_{\rm ev} h_0/D$ \cite{Routh2013}, where $h_0$ is the initial thickness of the film.  
If $\Pe<1$, the concentration gradient created by evaporation is quickly flattened by diffusion, and the colloid concentration remains uniform. 
On the other hand, if $\Pe>1$, the concentration gradient increases, and the colloids accumulate near the top of the film. 

If there are two types of colloids of different size \cite{LuoHui2008, Harris2009, Trueman2012, Atmuri2012}, the above consideration predicts that the larger colloids 
will accumulate near the free surface (\emph{large-on-top}), because larger colloids have a smaller diffusion constant, therefore a larger Peclet number. 
Recently, however, the opposite phenomenon has been reported by Fortini and coworkers \cite{Fortini2016}.  
By simulation and experiments, they have shown that smaller colloids appear on top of larger colloids (\emph{small-on-top}). 
They argued that this is due to the osmotic pressure of smaller colloids, but no quantitative theory has been given.

In this Letter, we show that the phenomenon can be explained by the standard diffusion model \cite{Routh2004} if the interaction between colloids are taken into account. 
We will use a simple hard sphere model, and show that the small-on-top structure is created by the cross-interaction between colloids of different sizes.
The effect of cross-interaction on colloidal motion is not symmetric: it is much stronger on larger colloids than smaller colloids and pushes the larger colloids towards the bottom of the film.
We will give a criterion when the small-on-top structure is created, and the corresponding experimental conditions, such as the drying rate, initial colloidal concentrations, and size ratio.

\begin{figure}[htbp]
  \centering
  \includegraphics[width=0.9\columnwidth]{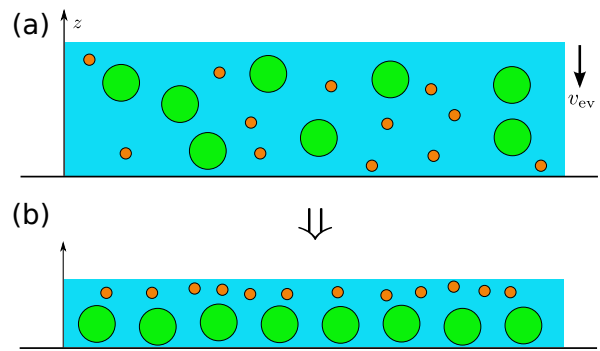}
  \caption{Drying of a binary colloidal solution in a film makes a 
    stratified film with small colloids on top of large colloids. }
  \label{fig:sketch}
\end{figure}

\emph{Evolution equations}. --We consider a thin film composed of two types of colloids of different size in solution (see Fig. \ref{fig:sketch}).  
In a thin film geometry, the lateral flow is not important and the film can be assumed to dry one-dimensionally.  
The evolution of the film height is $h(t)=h_0 - v_{\rm ev}t$, where $v_{\rm ev}$ is the evaporation rate.
The colloids are hard spheres with the radius $r_1$ and  $r_2$ (assuming $r_1<r_2$) and their volumes are $\nu_i=4\pi r_i^2/3$ ($i=1,2$). 
We define the size ratio by $\alpha=r_2/r_1>1$. 
The time-dependent volume fraction and number density are $\phi_i(z,t)$ and $n_i=\phi_i/\nu_i$, respectively.
Initially the colloidal solution is homogeneous with the volume fractions $\phi_i(z,0)=\phi_{0i}$.

For dilute hard-sphere mixture, the free energy density can be written as
\begin{equation}
  \label{eq:freeE}
  \frac{1}{k_BT} f(\phi_1,\phi_2) = \sum_i \frac{1}{\nu_i} \phi_i \ln\phi_i 
    + \sum_{i,j} \frac{1}{\nu_i \nu_j} a_{ij} \phi_i \phi_j
\end{equation}
where $a_{ij} = (2\pi/3) (r_i + r_j)^3$ is the second order virial coefficients for hard spheres. 
The chemical potential $\mu_i$ is then given by
\begin{equation}
  \label{eq:mu}
  \mu_i =  \frac{\partial f}{\partial n_i} = k_BT \left( \ln\phi_i + 1 
    + 2 \sum_j \frac{a_{ij}}{\nu_j} \phi_j \right) 
\end{equation}

The average velocity $v_i(z)$ of the colloids at $z$ is determined by the balance
of two forces. 
One is the thermodynamic force which is given by the gradient of the 
chemical potential (\ref{eq:mu}).  
The other is the hydrodynamic drag which is related to the colloid velocity $v_i$ by $\zeta_i v_i$, where $\zeta_i$ is the friction constant per colloid.  
The balance of these forces gives the average velocity
\begin{equation}
  \label{eq:vi}
  v_i = - \frac{1}{\zeta_i} \frac{\partial \mu_i}{\partial z} 
      = - \frac{D_i}{ k_BT} \frac{\partial \mu_i}{\partial z} .
\end{equation}
where we have used the Einstein relation $D_i=k_BT/\zeta_i$.
\change{In general, the diffusion constant takes a matrix form and depends on the colloidal concentrations due to direct and hydrodynamic interactions \cite{RSS, Dhont, DoiSoft}. 
Here we have only kept the diagonal terms and neglected the concentration-dependence.}

Given the velocity $v_i$, the time evolution of $\phi_i$ is obtained by the conservation equation
\begin{equation}
  \label{eq:conservation}
  \frac{\partial \phi_i}{\partial t} = - \frac{\partial \phi_i v_i}{\partial z}.
\end{equation}

Equations (\ref{eq:mu}), (\ref{eq:vi}), and (\ref{eq:conservation}) give 
\begin{equation}
  \frac{\partial \phi_i}{\partial t} = \frac{\partial}{\partial z} \Big[ \,\,
      \frac{\phi_i D_i}{k_BT} \frac{\partial \mu_i}{\partial z} \Big].
\end{equation}
Using the relation $r_2/r_1=\alpha$ and $\nu_2/\nu_1=\alpha^3$, the average velocities are explicitly written as
\begin{eqnarray}
  \label{eq:v1}
  v_1 &=& - D_1 \Big[ (\frac{1}{\phi_1} + 8) \frac{\partial \phi_1}{\partial z}  
    + (1+\frac{1}{\alpha})^3 \frac{\partial \phi_2}{\partial z} \Big] ,  \\ 
  \label{eq:v2}
  v_2 &=& - D_2 \Big[ \left(1+\alpha\right)^3 \frac{\partial \phi_1}{\partial z}
    + (\frac{1}{\phi_2} + 8) \frac{\partial \phi_2}{\partial z} \Big] .
\end{eqnarray}
The time evolution equations are 
\begin{eqnarray}
  \label{eq:diff1}
  \frac{\partial \phi_1}{\partial t} &=& D_1 \frac{\partial}{\partial z} 
    \Big[ (1 + 8\phi_1) \frac{\partial \phi_1}{\partial z}  
    + (1+\frac{1}{\alpha})^3 \phi_1 \frac{\partial \phi_2}{\partial z} \Big] \\ 
  \label{eq:diff2}
  \frac{\partial \phi_2}{\partial t} &=& D_2 \frac{\partial}{\partial z}
      \Big[ \left(1+\alpha \right)^3 \phi_2 
      \frac{\partial \phi_1}{\partial z}
    + (1 + 8\phi_2) \frac{\partial \phi_2}{\partial z} \Big]
\end{eqnarray}
These are coupled diffusion equations.
\change{They can also be derived from Onsager principle \cite{DoiSoft,appendix}.}
The boundary conditions at the substrate $z=0$ are $v_1 = v_2 = 0$.
At the free surface $z=h$, $v_1 = v_2  = -v_{\rm ev}$.

The coupled diffusion equations (\ref{eq:diff1}) and (\ref{eq:diff2}) can be 
made dimensionless by scaling the length to the initial film thickness $h_0$ 
and the time to the evaporation time scale
$\tau=h_0/v_{\rm ev}$ \cite{appendix}. 
This procedure introduces two Peclet numbers
\begin{equation}
  \Pe_1 = \frac{v_{\rm ev} h_0}{D_1}, \quad \Pe_2 = \frac{v_{\rm ev} h_0}{D_2} = \alpha \Pe_1 .
\end{equation}
\change{Here we have used Stokes-Einstein relation $D_i=k_BT/6\pi\eta r_i$, where $\eta$ is the fluid viscosity.}

\begin{figure*}[htbp]
  \centering
  \includegraphics[width=0.7\textwidth]{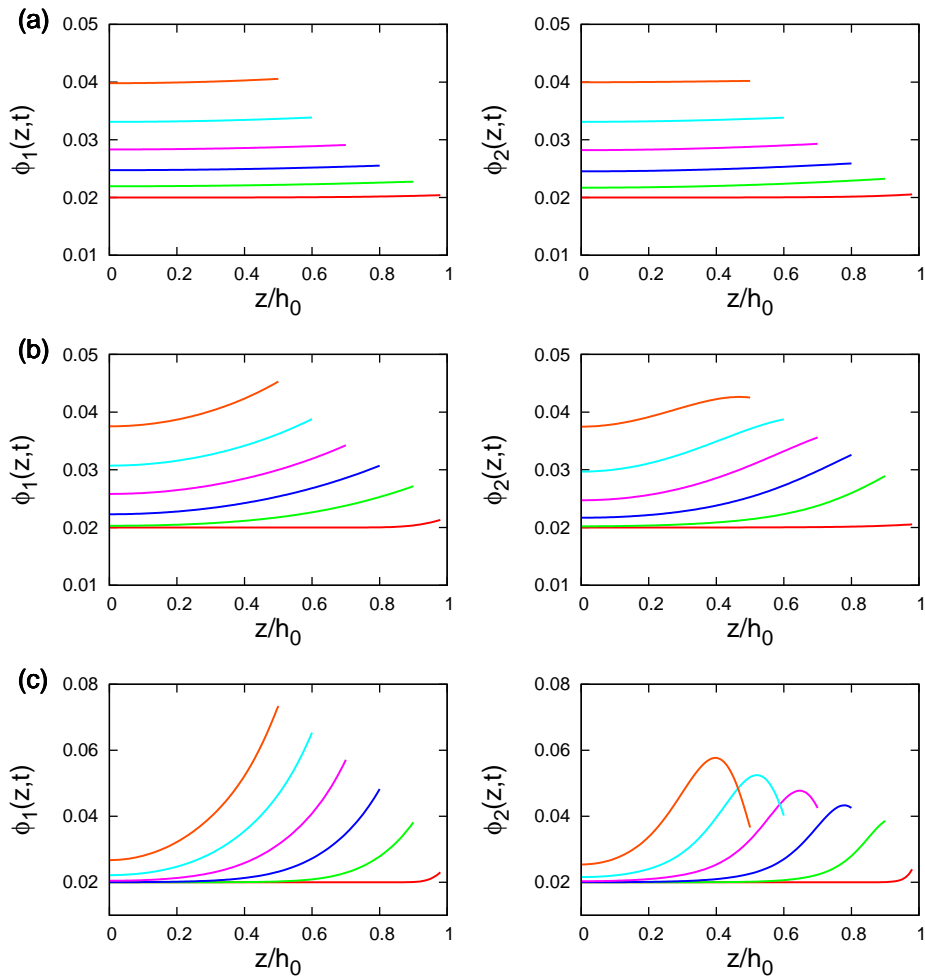} 
  \caption{Time variation of concentration profile of colloidal particles
having size ratio  \change{$\alpha=4$, (a) $\Pe_1=0.1$, $\Pe_2=0.4$. (b) $\Pe_1=1$, $\Pe_2=4$. (c) $\Pe_1=5$, $\Pe_2=20$.} The initial concentrations are $\phi_{01}=\phi_{02}=0.02$. The curves from bottom to top correspond to time $\tau=0.005, 0.1, 0.2, 0.3, 0.4, 0.5$.}
  \label{fig:num}
\end{figure*}

We solved the coupled diffusion equations numerically.
Figure \ref{fig:num} shows the representative concentration profiles at various times. 

When both Peclet numbers are less than 1 [Fig. \ref{fig:num}(a)], the perturbation due to the evaporation is small, and the concentration profiles for both colloids remain almost uniform, with slightly increase near the free surface.  
When both Peclet numbers are greater than 1 [Fig. \ref{fig:num}(c)], 
the free surface recedes faster than the diffusion, and the concentration becomes non-uniform. 
Initially both colloids accumulated at the free surface, but at later times, the concentration gradient of the smaller colloid becomes large, and eventually drives the big colloids to the bottom. 
Figure \ref{fig:num}(b) shows the intermediate state, where the concentration profile of large colloid near the free surface becomes flat at late times, but a clear stratification has not fully developed yet.


\emph{Analytic theory.} --We can understand the mechanism by taking a close look at the average velocities Eqs. (\ref{eq:v1}) and (\ref{eq:v2}).  
If there is no interaction between the colloids, the equations take a simple form
\begin{equation}
  \label{eq:diff_ni}
  v_i = - D_i \Big[ \frac{1}{\phi_i} \frac{\partial \phi_i}{\partial z} \Big],
\end{equation}
which gives a pair of uncoupled diffusion equations
\begin{equation}
  \frac{\partial \phi_i}{\partial t} = D_i \frac{\partial^2 \phi_i}{\partial z^2}.
\end{equation}

In Eqs. (\ref{eq:diff1}) and (\ref{eq:diff2}), the terms $8 (\partial \phi_i/\partial z)$ terms come from the self-interaction (virial coefficient $a_{ii}$), while the $(1+1/\alpha)^3 (\partial \phi_2/\partial z)$ and $(1+\alpha)^3 (\partial \phi_1/\partial z)$ terms originate from the cross-interaction (virial coefficients $a_{12}=a_{21}$). 
One can immediately see that the cross-interaction term affects the larger colloids much more strongly than the smaller colloids due to the factor of $(1+\alpha)^3$. 

The small-on-top structure forms when the first term in Eq.~(\ref{eq:v2}) becomes larger than the second term, either due to a large size ratio $\alpha$, or a strong concentration gradient of smaller colloids $\partial \phi_1/\partial z$.
In this case, the larger colloids are driven to the substrate 
while smaller colloids are left near the top surface. 
The condition for this phenomenon to happen can be written as 
\begin{equation}
  \label{eq:layering1}
  (1+\alpha)^3 \frac{\partial \phi_1}{\partial z} > C 
  \frac{1}{\phi_2} \frac{\partial \phi_2}{\partial z},
\end{equation}
where $C$ is a factor which can be regarded as a fitting parameter in our model.
Since our theory accounts for the effect of interaction up to the second order term,
we expect the condition (\ref{eq:layering1}) to be valid at dilute regime.
At late times, the stratified structure formed at low concentration would persist over to higher concentrations and the final film remains small-on-top structure.

We can write the condition (\ref{eq:layering1}) in terms of experimental parameters. 
We use the results for non-interacting colloids from Eq.~(\ref{eq:diff_ni}) as a first-order approximation.  
\change{The evolution of a drying film with one type of colloids \cite{Routh2004, Style2011} or polymers \cite{Tsige2004, Okuzono2006, LuoLing2016} has been studied. 
In Ref. \cite{Okuzono2006}, the same diffusion model was used and analytic results are derived at the surface,}
\begin{eqnarray}
  \label{eq:dpdzh}
  \frac{\partial \phi_i}{\partial z} &=& \frac{v_{\rm ev}}{D_i} \phi_{hi} \\
  \label{eq:phih}
  \phi_{hi} &\approx& \left( 1 + \sqrt{ \frac{4 v_{\rm ev}^2}{\pi D_i} } t^{1/2} \right) \phi_{0i} \approx ( 1+\Pe_i ) \phi_{0i} 
\end{eqnarray}
In the second equation, we have used the characteristic time $t=h_0^2/D_i$. 

The condition for the small-on-top structure (\ref{eq:layering1}) is then simplified 
\begin{eqnarray}
  && (1+\alpha)^3 \frac{v_{\rm ev}}{D_1} \phi_{h1} > 
  C \frac{1}{\phi_2} \frac{v_{\rm ev}}{D_2} \phi_{h2} \nonumber \\
  & \Rightarrow & (1+\alpha)^3 \frac{D_2}{D_1} \phi_{h1} > C .
  \label{eq:layering2a}
\end{eqnarray}
Hence for large $\alpha$, the condition is
\begin{equation}
  \label{eq:layering2}
  \alpha^2 ( 1 + \Pe_1 ) \phi_{01} > C .
\end{equation}

It is interesting to note that the condition (\ref{eq:layering2}) does not depend on $\phi_{02}$. 
This is plausible because the cross-interaction term in Eq.~(\ref{eq:v2}), 
which is responsible for driving the big colloids to the bottom, does not depend on $\phi_{02}$.
The size ratio comes in term of $\alpha^2$ in (\ref{eq:layering2}), indicating that the size asymmetry has a strong effect on the stratification.


\emph{State diagrams.} --To test our analytic formula, we solved the coupled diffusion equations (\ref{eq:diff1}) and (\ref{eq:diff2}) for large sets of parameters $(\Pe_1, \alpha, \phi_{01}, \phi_{02})$. 
We stopped the numerical calculation when $h=h_0/2$ and regarded the structure at this state as the indicative of the final structure.  
We did this because our model ceases to be valid at high concentration and whether the system takes the stratified structure or not can be discussed at this state.

\begin{figure}[htbp]
  \centering
  \includegraphics[width=1.0\columnwidth]{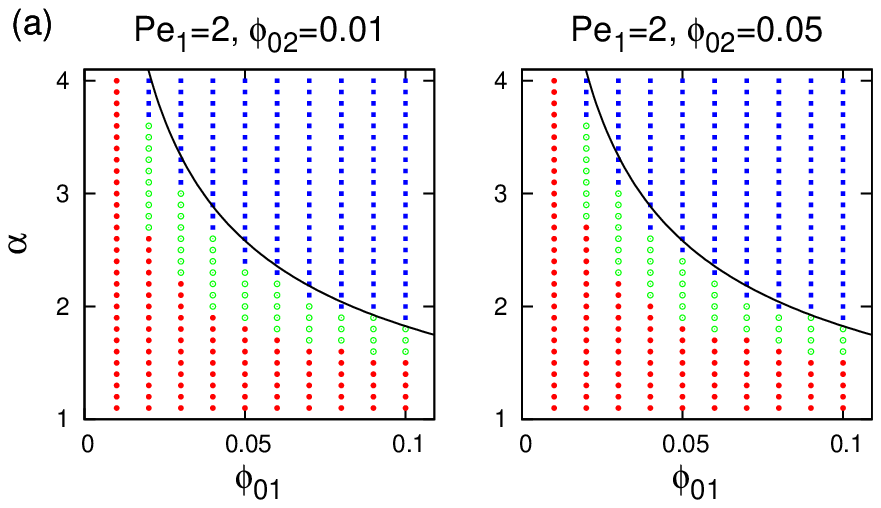}
  \includegraphics[width=1.0\columnwidth]{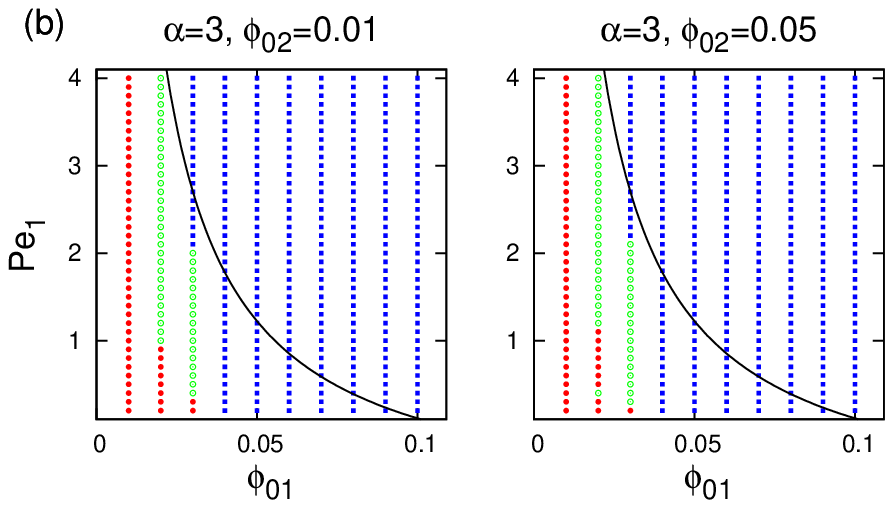}
  \caption{(a) State diagrams in the $\phi_{01}$-$\alpha$ plane. Parameters are $\Pe_1=2$, $\phi_{02}=0.01$ (left) and $\phi_{02}=0.05$ (right). (b) State diagrams in the $\phi_{01}$-$\Pe_1$ plane. Parameters are $\alpha=3$, $\phi_{02}=0.01$ (left) and $\phi_{02}=0.05$ (right). The solid curve corresponds to $\alpha^2(1+\Pe_1)\phi_{10}=1$. }
  \label{fig:state}
\end{figure}

We extrapolated the concentration profile at the last step of the calculation, and constructed an expected state diagram of the dried state. 
We judged the dried state will have small-on-top structure if there is a peak of $\phi_2$, \emph{i.e.}, if $\partial \phi_2/\partial z |_{z=h}$ is negative at the last step of the calculation.
We used blue squares (\textcolor{blue}{$\scriptscriptstyle \blacksquare$}) to indicate these states.   
\change{If $\partial \phi_2/\partial z |_{z=h}$ is positive, and for some value of $z$ in the range of $0<z<h$, $\phi_2(z)$ has a negative curvature (\emph{i.e.}, $\partial^2 \phi_2/\partial z^2 <0$), the small-on-top structure may form at late times.}
Therefore we classified the state intermediate (\textcolor{green}{$\textstyle \circ$}).
Otherwise, the dried state will have either the large-on-top structure, or almost homogeneous distributions of both smaller and larger colloids.
We labeled these states using the symbol (\textcolor{red}{$\bullet$}). 
These states are shown in Fig. \ref{fig:state}.

Figure \ref{fig:state}(a) shows the results in the $\phi_{01}$-$\alpha$ plane for $\Pe_1=2$ and initial concentrations $\phi_{02}=0.01, 0.05$. 
For these two different starting concentrations, the state diagrams are similar, confirming our expectation that the state is independent of $\phi_{20}$. 
\change{In the parameter range we considered, the small-on-top structure appears when either the size ratio is large, or the initial concentration $\phi_{01}$ is large, which eventually results in a large concentration gradient $\partial \phi_1/\partial z$.} 
Both factors produce a large cross-interaction term which drives the larger colloids to the bottom.  
The solid curve in Fig. \ref{fig:state}(a) corresponds to Eq.~(\ref{eq:layering2}) with $C=1$, which identifies the boundary of small-on-top structure rather well. 

Figure \ref{fig:state}(b) shows the results in the $\phi_{01}$-$\Pe_1$ plane for the size ratio \change{$\alpha=3$} and initial concentrations $\phi_{02}=0.01, 0.05$. 
Again the theoretical curve qualitatively explains the boundary of small-on-top region.
One should note even at $\Pe_1<1$, there are noticeable parameter space ($\phi_{01}>0.05$) where small-on-top structure appears. 

\begin{figure}[htbp]
  \centering
  \includegraphics[width=1.0\columnwidth]{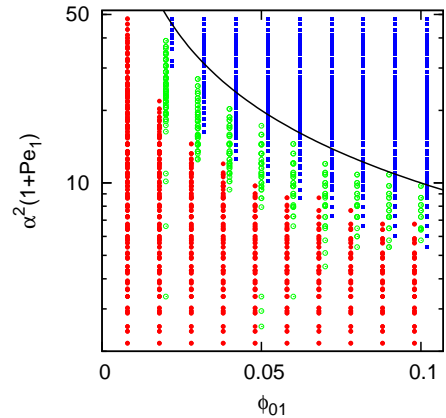}
  \caption{Master plot of the state diagram in the $\phi_{01}$--$\alpha^2(1+\Pe_1)$ plane. The solid curve corresponds to $\alpha^2(1+\Pe_1)\phi_{01}=1$. \change{The states labeled by (\textcolor{blue}{$\scriptscriptstyle \blacksquare$}) and (\textcolor{red}{$\bullet$}) are shifted slightly in the $\phi_{01}$-axis for a better view.}}
  \label{fig:master}
\end{figure}

Figure \ref{fig:master} is a master plot collecting all numerical results, where the vertical axis is taken to be $\alpha^2(1+\Pe_1)$.  
The agreement between the theoretical prediction and numerical results is not perfect,
but Eq.~(\ref{eq:layering2}) has captured the general trend of the state boundary.

\emph{Discussion and conclusion.} --If there is no interaction between colloids, the larger colloids will accumulate near the surface when $\Pe_2>1$.
The condition for this to happen is simply
\begin{equation}
  \label{eq:skin}
  \Pe_2 > 1, \quad \textrm{or} \quad \Pe_1 > 1/\alpha.
\end{equation}

\begin{figure}[htbp]
  \centering
  \includegraphics[width=0.9\columnwidth]{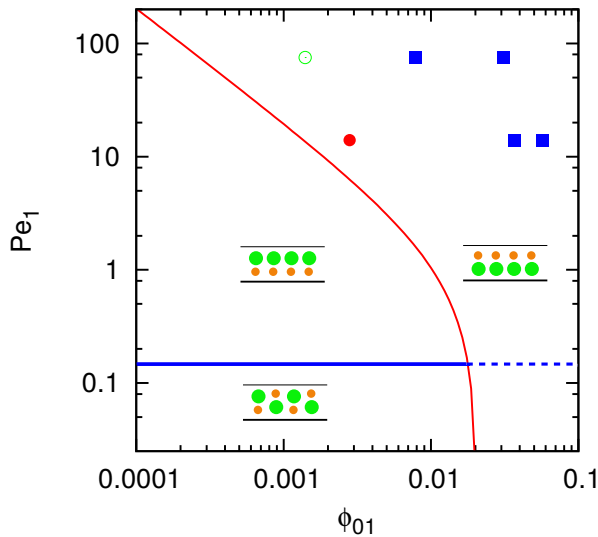}
  \caption{State diagram when close-packing is also considered. The size ratio is $\alpha=7$. The symbols are results taken from Ref. \cite{Fortini2016}. The labeling is the same as in Fig.~\ref{fig:state} and Fig.~\ref{fig:master} and discussed in \cite{appendix}.}
  \label{fig:state_dia}
\end{figure}

Equation (\ref{eq:skin}) is plotted as the blue line in Fig. \ref{fig:state_dia} for $\alpha=7$. 
Above this line, at late times the larger colloids reach the close-packing earlier than the smaller colloids and form the top layer.
However, at early time, the concentration gradient of the smaller colloids, combined with a large size-asymmetry, results a large cross-interaction term which drives the larger colloids to the bottom. 
This condition is given by Eq.~(\ref{eq:layering2}) and is plotted as a red curve in Fig.~\ref{fig:state_dia}.
On the right-hand side of the red curve, the accumulation of larger colloids near the free surface is preempted by the cross-interaction at early times.

\change{In Fig.~\ref{fig:state_dia}, we also compare our results with the simulation and experimental results of Ref.~ \cite{Fortini2016}, shown in symbols \cite{appendix}. 
The overall agreement is good except one experimental data point (the red circle), which also appears to be closest to the transition line.}


\change{Extension and improvements can be made in our simple diffusion model.
Besides the binary mixture of colloidal particles, mixture of polymers and nanoparticles is another interesting system \cite{Jouault2014, ChengShengfeng2016}. 
Our theory may shed light on the fabrication of polymer nanocomposite by film drying.
The limitation of dilute solution can also be removed by using a more general equation of state \cite{Carnahan1969, Hansen-Goos2006}. 
We have used a simple hard-sphere model, while various types of interaction between colloids can be introduced through the second-order virial coefficient.
In our model we also neglect the effect of hydrodynamic interactions. 
This can be amended by using a concentration-dependent diffusion constant to replace the Stokes value. 
Nevertheless, we emphasis that the phenomenon described here is quite robust and happens at low concentrations, in the region where our simple diffusion model would be sufficient.}
  
In summary, we have implemented a diffusion model for the drying colloidal mixtures which incorporates explicitly the interaction between different colloid types. 
The smaller colloids exclude the larger colloids and accumulate near the free surface, which stems from the cross-interactions. 
The cross-interactions depend on the concentration gradient of the smaller colloids and the large-to-small colloid size ratio. 
This is a purely out-of-equilibrium phenomenon because the concentration gradient is driven by the evaporation.  
It also happens at low concentrations, in the region where the diffusion model would be sufficient.

This work was supported by the National Natural Science Foundation of China (NSFC) through the Grant No. 21434001, 21504004, 21574006, and 21622401. M. D. acknowledges the financial support of the Chinese Central Government in the Thousand Talents Program. 

\bibliography{colloids}



\end{document}